\documentclass[conference,10pt]{IEEEtran}
\usepackage{amsmath,amssymb,amsmath,amsfonts}
\usepackage{bm}
\usepackage{bbm}
\usepackage{graphicx}
\usepackage[noadjust]{cite}
\usepackage{citesort}
\usepackage{epstopdf}
\usepackage{algorithmic}
\usepackage{url}
\usepackage{makecell}

\usepackage{gensymb}
\usepackage{color}
\usepackage{lipsum}
\usepackage{float}
\usepackage{epstopdf}
\usepackage[mathcal]{eucal}
\usepackage{subfiles}
\usepackage[T1]{fontenc}
\usepackage{siunitx}
\usepackage{tabulary}

\usepackage{array}
\usepackage{adjustbox}
\usepackage[dvipsnames]{xcolor}
\usepackage{times}
\usepackage{placeins}
\usepackage{textcomp}
\usepackage[font=small]{caption}
\usepackage{xcolor}
\usepackage{tabularx}
\usepackage{bm}
\usepackage{enumerate}
\usepackage{tikz}
\usepackage{pgfplots}
\usepackage{epstopdf}
\usetikzlibrary{shapes,arrows}
\usepackage[utf8]{inputenc}
\usepackage{mathtools}
\usepackage[utf8]{inputenc}
\usepackage{xcolor}
\usepackage{tikz}
\usetikzlibrary{chains,arrows,calc,positioning}

\pgfplotsset{compat=1.16}
\definecolor{bittersweet}{rgb}{1.0, 0.44, 0.37}
\definecolor{glaucous}{rgb}{0.38, 0.51, 0.71}
\definecolor{gainsboro}{rgb}{0.86, 0.86, 0.86}
\definecolor{babyblueeyes}{rgb}{0.63, 0.79, 0.95}
\definecolor{silver}{rgb}{0.75, 0.75, 0.75}
\definecolor{neoncarrot}{rgb}{1.0, 0.64, 0.26}

\usepackage{soul}
\usepackage{algorithm}
\usepackage{varwidth}
\usepackage{breqn}
\usepackage{booktabs}
\usepackage{multirow}

\usepackage{enumitem}
\usepackage{tabulary}

\usepackage[normalem]{ulem}

\usepackage[acronyms,nonumberlist,nopostdot,nomain,nogroupskip]{glossaries}
\newacronym{quic}{QUIC}{Quick UDP Internet Connections}
\newacronym{3gpp}{3GPP}{3rd Generation Partnership Project}
\newacronym{adc}{ADC}{Analog to Digital Converter}
\newacronym{5g}{5G}{5th generation}
\newacronym{aimd}{AIMD}{Additive Increase Multiplicative Decrease}
\newacronym{am}{AM}{Acknowledged Mode}
\newacronym{amc}{AMC}{Adaptive Modulation and Coding}
\newacronym{aqm}{AQM}{Active Queue Management}
\newacronym{awgn}{AGWN}{Additive White Gaussian Noise}
\newacronym{afd}{AFD}{Austin Fire Department}
\newacronym{balia}{BALIA}{Balanced Link Adaptation}
\newacronym{bdp}{BDP}{Bandwidth-Delay Product}
\newacronym{bf}{BF}{Beamforming}
\newacronym{cc}{CC}{Congestion Control}
\newacronym{cdf}{CDF}{Cumulative Distribution Function}
\newacronym{cn}{CN}{Core Network}
\newacronym{cqi}{CQI}{Channel Quality Information}
\newacronym{cp}{CP}{Control Plane}
\newacronym{csirs}{CSI-RS}{Channel State Information - Reference Signal}
\newacronym{dc}{DC}{Dual Connectivity}
\newacronym{dce}{DCE}{Direct Code Execution}
\newacronym{dci}{DCI}{Downlink Control Information}
\newacronym{dl}{DL}{Downlink}
\newacronym{dmr}{DMR}{Deadline Miss Ratio}
\newacronym{dmrs}{DMRS}{DeModulation Reference Signal}
\newacronym{e2e}{E2E}{End-to-End}
\newacronym{ecn}{ECN}{Explicit Congestion Notification}
\newacronym{edf}{EDF}{Earliest Deadline First}
\newacronym{enb}{eNB}{evolved Node Base}
\newacronym{epc}{EPC}{Evolved Packet Core}
\newacronym{es}{ES}{Edge Server}
\newacronym{fdma}{FDMA}{Frequency Division Multiple Access}
\newacronym{fdd}{FDD}{Frequency Division Duplexing}
\newacronym[firstplural=Radio Access Technologies (RATs)]{rat}{RAT}{Radio Access Technology}
\newacronym{fs}{FS}{Fast Switching}
\newacronym{ftp}{FTP}{File Transfer Protocol}
\newacronym{gnb}{gNB}{Next Generation Node Base}
\newacronym{harq}{HARQ}{Hybrid Automatic Repeat reQuest}
\newacronym{hetnet}{HetNet}{Heterogeneous Network}
\newacronym{hh}{HH}{Hard Handover}
\newacronym{hol}{HOL}{Head-of-Line}
\newacronym{ia}{IA}{Initial Access}
\newacronym{imt}{IMT}{International Mobile Telecommunication}
\newacronym{iot}{IoT}{Internet of Things}
\newacronym{los}{LOS}{Line of Sight}
\newacronym{lte}{LTE}{Long Term Evolution}
\newacronym{m2m}{M2M}{Machine to Machine}
\newacronym{mac}{MAC}{Medium Access Control}
\newacronym{mc}{MC}{Multi-Connectivity}
\newacronym{mcs}{MCS}{Modulation and Coding Scheme}
\newacronym{mec}{MEC}{Mobile Edge Cloud}
\newacronym{mi}{MI}{Mutual Information}
\newacronym{mimo}{MIMO}{Multiple Input, Multiple Output}
\newacronym{mmwave}{mmWave}{millimeter wave}
\newacronym{mr}{MR}{Maximum Rate}
\newacronym{mss}{MSS}{Maximum Segment Size}
\newacronym{mtd}{MTD}{Machine-Type Device}
\newacronym{mtu}{MTU}{Maximum Transmission Unit}
\newacronym{nfv}{NFV}{Network Function Virtualization}
\newacronym{nlos}{NLOS}{Non Line of Sight}
\newacronym{nr}{NR}{New Radio}
\newacronym{ofdm}{OFDM}{Orthogonal Frequency Division Multiplexing}
\newacronym{pdcch}{PDCCH}{Physical Downlonk Control Channel}
\newacronym{pdcp}{PDCP}{Packet Data Convergence Protocol}
\newacronym{pdsch}{PDSCH}{Physical Downlink Shared Channel}
\newacronym{pdu}{PDU}{Packet Data Unit}
\newacronym{pf}{PF}{Proportional Fair}
\newacronym{pgw}{PGW}{Packet Gateway}
\newacronym{phy}{PHY}{Physical}
\newacronym{pbch}{PBCH}{Physical Broadcast Channel}
\newacronym[plural=\gls{mme}s,firstplural=Mobility Management Entities (MMEs)]{mme}{MME}{Mobility Management Entity}
\newacronym{prb}{PRB}{Physical Resource Block}
\newacronym{pss}{PSS}{Primary Synchronization Signal}
\newacronym{pucch}{PUCCH}{Physical Uplink Control Channel}
\newacronym{pusch}{PUSCH}{Physical Uplink Shared Channel}
\newacronym{rach}{RACH}{Random Access Channel}
\newacronym{ran}{RAN}{Radio Access Network}
\newacronym{red}{RED}{Robotics Emergency Deployment}
\newacronym{rf}{RF}{Radio Frequency}
\newacronym{rlc}{RLC}{Radio Link Control}
\newacronym{rlf}{RLF}{Radio Link Failure}
\newacronym{rrc}{RRC}{Radio Resource Control}
\newacronym{rrm}{RRM}{Radio Resource Management}
\newacronym{rr}{RR}{Round Robin}
\newacronym{rs}{RS}{Remote Server}
\newacronym{rsrp}{RSRP}{Reference Signal Received Power}
\newacronym{rss}{RSS}{Received Signal Strength}
\newacronym{rtt}{RTT}{Round Trip Time}
\newacronym{rw}{RW}{Receive Window}
\newacronym{rx}{RX}{Receiver}
\newacronym{sa}{SA}{standalone}
\newacronym{sack}{SACK}{Selective Acknowledgment}
\newacronym{sap}{SAP}{Service Access Point}
\newacronym{sch}{SCH}{Secondary Cell Handover}
\newacronym{scoot}{SCOOT}{Split Cycle Offset Optimization Technique}
\newacronym{sdma}{SDMA}{Spatial Division Multiple Access}
\newacronym{sinr}{SINR}{Signal to Interference plus Noise Ratio}
\newacronym{sm}{SM}{Saturation Mode}
\newacronym{snr}{SNR}{Signal to Noise Ratio}
\newacronym{son}{SON}{Self-Organizing Network}
\newacronym{ss}{SS}{Synchronization Signal}
\newacronym{srs}{SRS}{Sounding Reference Signal}
\newacronym{sss}{SSS}{Secondary Synchronization Signal}
\newacronym{tb}{TB}{Transport Block}
\newacronym{tcp}{TCP}{Transmission Control Protocol}
\newacronym{tdd}{TDD}{Time Division Duplexing}
\newacronym{tdma}{TDMA}{Time Division Multiple Access}
\newacronym{tfl}{TfL}{Transport for London}
\newacronym{tm}{TM}{Transparent Mode}
\newacronym{trp}{TRP}{Transmitter Receiver Pair}
\newacronym{tti}{TTI}{Transmission Time Interval}
\newacronym{ttt}{TTT}{Time-to-Trigger}
\newacronym{tx}{TX}{Transmitter}
\newacronym{ue}{UE}{User Equipment}
\newacronym{ul}{UL}{Uplink}
\newacronym{uml}{UML}{Unified Modeling Language}
\newacronym{um}{UM}{Unacknowledged Mode}
\newacronym{utc}{UTC}{Urban Traffic Control}
\newacronym{vm}{VM}{Virtual Machine}
\newacronym{rsrq}{RSRQ}{Reference Signal Received Quality}
\newacronym{rssi}{RSSI}{Received Signal Strength Indicator}
\newacronym{crs}{CRS}{Cell Reference Signal}
\newacronym{comp}{CoMP}{Coordinated Multi-Point}
\newacronym{cran}{C-RAN}{Cloud \acrlong{ran}}
\newacronym{ca}{CA}{Carrier Aggregation}
\newacronym{cco}{CC}{Carrier Component}
\newacronym{nsa}{NSA}{Non Stand Alone}
\newacronym{embb}{eMBB}{Enhanced Mobility Broadband}
\newacronym{bsr}{BSR}{Buffer Status Report}
\newacronym{srb}{SRB}{Service Radio Bearer}
\newacronym{scm}{SCM}{Spatial Channel Model}
\newacronym{sctp}{SCTP}{Stream Control Transmission Protocol}
\newacronym{mptcp}{MPTCP}{Multi-path TCP}
\newacronym{ietf}{IETF}{Internet Engineering Task Force}
\newacronym{os}{OS}{Operating System}
\newacronym{tls}{TLS}{Transport Layer Security}
\newacronym{rfc}{RFC}{Request for Comments}
\newacronym{http}{HTTP}{HyperText Transfer Protocol}
\newacronym{nat}{NAT}{Network Address Translation}
\newacronym{api}{API}{Application Programming Interface}
\newacronym{rto}{RTO}{Retransmission Timeout}
\newacronym{psc}{PSC}{Public Safety Communication}
\newacronym{rpgm}{RPGM}{Reference Point Group Mobility}
\newacronym{ic}{IC}{Incident Command}
\newacronym{rsu}{RSU}{Road Side Unit}
\newacronym{uav}{UAV}{unmanned aerial vehicle}
\newacronym{usv}{USV}{Unmanned Surface Vehicle}
\newacronym{uas}{UAS}{Unmanned Aerial System}
\newacronym{iab}{IAB}{Integrated Access and Backhaul}
\newacronym{qoe}{QoE}{Quality of Experience}
\newacronym{ssim}{SSIM}{Structural Similarity Index}
\newacronym{psnr}{PSNR}{Peak Signal to Noise Ratio}
\newacronym{bs}{BS}{Base Station}
\newacronym{mu}{MU}{Multiple User}
\newacronym{ag}{AG}{Air-to-Ground}
\newacronym{af}{AF}{Array Factor}
\newacronym{ula}{ULA}{Uniform Linear Array}
\newacronym{upa}{UPA}{Uniform Planar Array}
\newacronym{lcs}{LCS}{Local Coordinate System}
\newacronym{psd}{PSD}{Power Spectral Density}
\newacronym{vq}{VQ}{vector quantization}
\newacronym{a2g}{A2G}{air-to-ground}
\newacronym{em}{EM}{electromagnetic}
\newacronym{vae}{VAE}{variational autoencoder}

\makeatletter

 \let\oldforeign@language\foreign@language
 \DeclareRobustCommand{\foreign@language}[1]{%
   \lowercase{\oldforeign@language{#1}}}

\usepackage[font=small]{caption}

\usepackage{dblfloatfix}

\def\nb0{{\mathbf{0}}}
\def\nb1{{\mathbf{1}}}

\makeatother

\IEEEoverridecommandlockouts

\newcommand{\norm}[1]{\left\lVert#1\right\rVert}

\usepackage[top=0.705in, left=0.625in, bottom=1in, right=0.625in]{geometry}

\begin{document}
%
\title{A Preliminary Assessment of Midhaul Links \\ at 140 GHz using Ray-Tracing}
%

\author{
%
\IEEEauthorblockN{Sravan Reddy Chintareddy$^{\sharp}$ 
\quad Marco Mezzavilla$^{\dagger}$ \quad Sundeep Rangan$^{\dagger}$ \quad  Morteza Hashemi$^{\sharp}$} 
\IEEEauthorblockA{$^{\sharp}$The University of Kansas, Lawrence, KS, USA}
\IEEEauthorblockA{$^{\dagger}$NYU Tandon School of Engineering, Brooklyn, NY, USA}
\thanks{M. Mezzavilla and S. Rangan were supported by
NSF grants  1302336,  1564142,  1547332, and 1824434,  NIST, SRC, and the industrial affiliates of NYU WIRELESS. Sravan R. Chintareddy and M. Hashemi were supported by NSF grants CNS-1948511 and CNS-1955561.}
 \vspace{-6mm}
}

\maketitle

\begin{abstract}
 The ever-growing demand for mobile data necessitates a transport network architecture that can withstand the 5G-and-beyond multi-Gbps traffic requirements. To cater for such unprecedented demand, studies are being conducted to incorporate TeraHertz (THz) communications in future mobile networks. In this paper, we consider an urban environment and evaluate the feasibility of THz wireless midhaul links for the transport networks between the Central Units (CU) and Distributed Units (DU) in a disaggregated 5G network architecture with functional splits.
 Our goal is to study the feasibility of midhaul links at 140 GHz by minimizing the number of required CUs to serve all the DUs. To this end, we define several policies for selecting CU and DU nodes in order to determine the peak data rate that can be supported over each link between a CU and DU. Our numerical results based on ray-tracing suggest that wireless links at 140 GHz with 3GPP option 2 as High Layer Split (HLS) represents a promising technology for midhaul transport networks.
\end{abstract}


\IEEEpeerreviewmaketitle

\section{Introduction}
\label{sec:intro}

\begin{figure}
 	\centering
	\includegraphics[width=\linewidth]{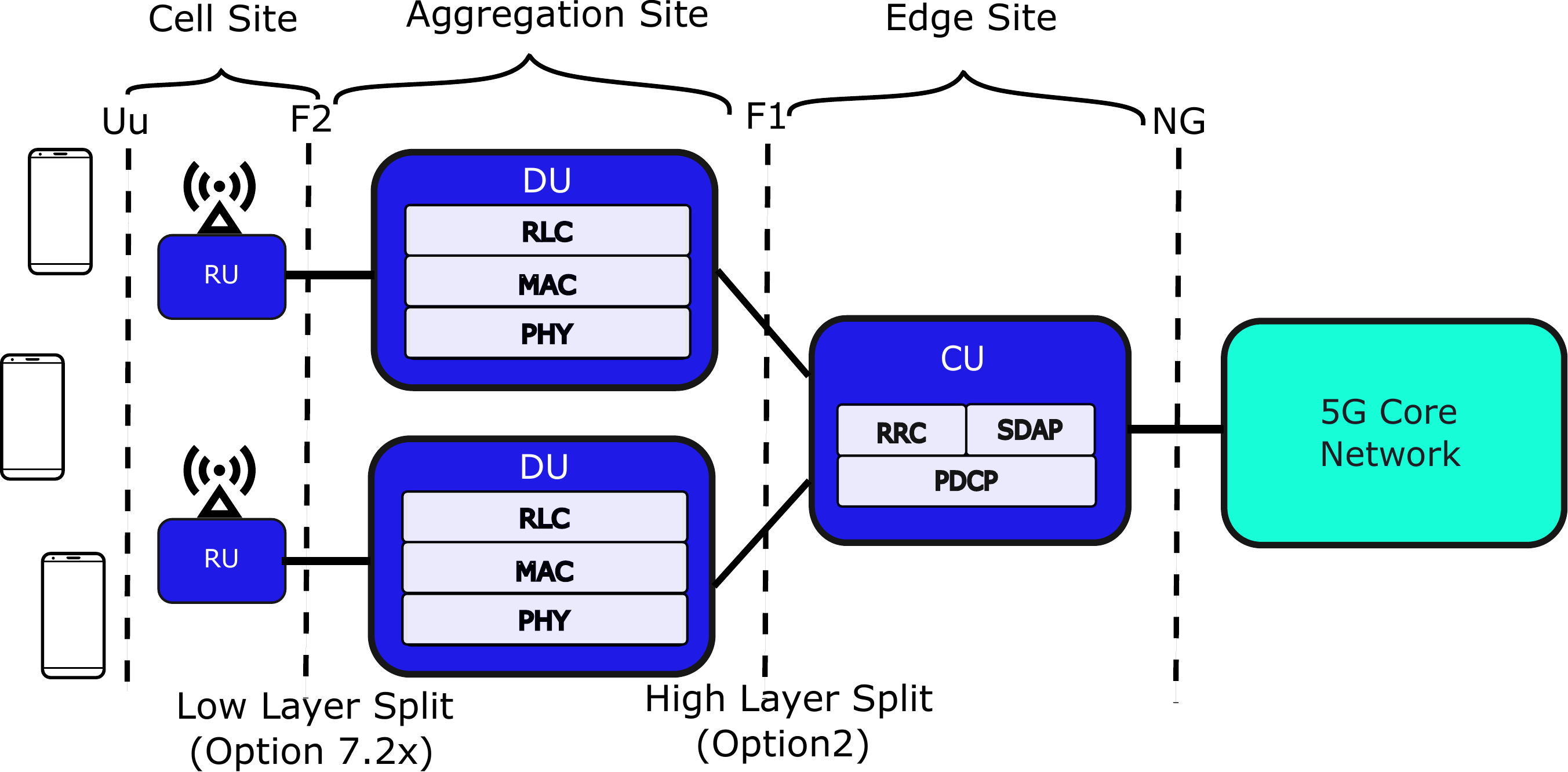}
	\caption{{5G NR architecture with functional split and interfaces.}}
	\label{fig:RAN}
 	\vspace{-4mm}
\end{figure}

 The massive volume of data traffic enabled by 5G New Radio (5G-NR) leads to an extreme capacity requirements on the transport network. In order to withstand this extreme demand on the transport network, 3GPP identified functional split as a potential solution~\cite{3GPP}. A functional split dictates the amount of functions that should reside at the cell sites locally, and the amount of functions that should be centralized at more powerful data centers~\cite{larsen2018survey}. A number of different functional split options and interfaces are being investigated to be used in 5G-NR~\cite{NGMN,TS38300,TS38475} where the 5G base station (gNB) is divided into Remote Unit (RU), Distributed Unit (DU), and a Central Unit (CU), as shown in Fig.~\ref{fig:RAN}.

Moving forward to the sixth generation (6G), the overwhelming demand of data traffic on transport networks as well as requirements for multi-Gbps data rates will further continue~\cite{akyildiz2014terahertz}. Considering this growth of data traffic, integrating new chunks of untapped spectrum above 100 GHz is identified as a potential solution to achieve ultra high data rates in 6G networks~\cite{rappaport2019wireless}. Furthermore, as the wavelength is smaller at terahertz frequencies, many antenna elements can be packed into a small form factor, which enables Ultra-massive Multiple Input, Multiple Output (UM-MIMO) techniques~\cite{akyildiz2016realizing}. Higher available bandwidths and larger density of antenna elements together demand for more sophisticated signal processing techniques to enable multi-Gbps data rates. To this end, the authors in~\cite{sarieddeen2020overview} studied THz-specific signal processing techniques for wireless communications that include waveform design
and modulation, beamforming and precoding, index modulation, channel estimation, channel coding, and data detection. 

While the studies mentioned earlier explored the theoretical aspects of THz communications, field experiments are not performed due to lack of hardware. In the recent years, due to advancements in electronics and hardware availability, several outdoor experiments are being conducted at 140 GHz~\cite{simsek2018140,xing2018propagation} to study the feasibility of using 140 GHz in futuristic mobile networks. In particular, the authors in~\cite{nguyen2018comparing} identified that the signal characteristics like pathloss, delay spread, maximum number of strongest paths in 28 GHz and 140 GHz systems are similar, suggesting that the latter can be considered for wireless broadband applications. Also, several studies suggested that frequencies beyond 100 GHz can be used for point-to-point links~\cite{rappaport2019wireless,akyildiz2014terahertz}. However, to the best of our knowledge midhaul links between CU-DU at 140 GHz has not been fully explored yet. 
In this paper, we consider the design aspects of efficient midhaul links at 140 GHz between CU and DU in an urban environment. In particular, our work aims to minimize the  number of CUs that needs to be commissioned to serve all the DUs. To this end, we define optimal policies for selecting CU and DU nodes, and determine the peak data rate that can be supported over each link between CU and DUs. 
\section{Background and Related Work}
\label{sec:background}

\begin{figure}[t!]
    \includegraphics[width=\columnwidth]{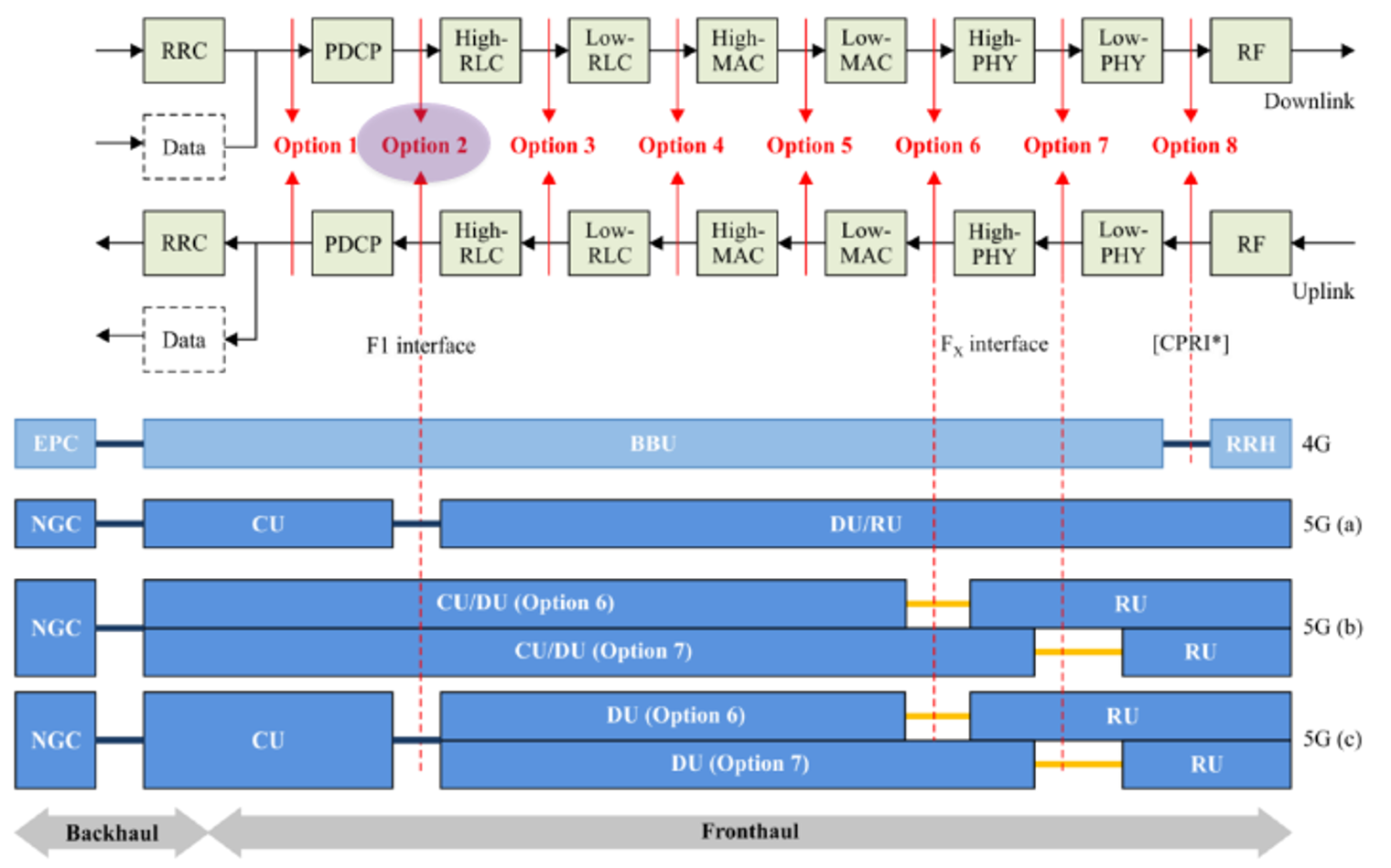}
    \caption{Mapping of CU/DU/RU functions for 5G (a) high layer split (F1), 5G (b) low layer split (Fx), and 5G (c) in a cascaded split architecture with different split options~\cite{5GTransport2018}.}
    \label{fig:cascaded}
    \vspace{-5mm}
\end{figure}

As discussed in section~\ref{sec:intro}, the overall goal of the functional split is to determine the amount of computation that can be carried out at the cell site and remote data centers. For instance,  Fig.~\ref{fig:cascaded} shows a cascaded view of functional split where eight split points and several sub-options have been proposed by 3GPP~\cite{3GPP}. In this figure, the interfaces on fronthaul and midhaul based on the 3GPP options are shown as dotted lines. Different 5G options, namely 5G (a), 5G (b), and 5G (c) are also shown in this cascaded view. Options 5G (a) and 5G (b) have a single split point that divides the gNB into CU and DU, while 5G (c) considers both HLS and LLS and the gNB is split into CU, DU, and RU. 
In this context, the authors in~\cite{Thor,Terranova,EuWireless} provide an estimation on the demand of mobile data on transport networks when considering the 3GPP split options. However, a cascaded view of functional split, using both HLS and LLS, along with an evaluation of midhaul wireless links, are not fully explored.  

In addition to the functional splits as a potential solution, it should be noted that 5G-and-beyond networks with small cell sizes require more base stations to provide ubiquitous coverage~\cite{bhushan2014network}. Hence, using optical fiber would be a costly option for such dense networks. To address this problem, integrated access and backhaul (IAB) was proposed by 3GPP ~\cite{zhang2021survey}. 
However, thus far the research on IAB networks  are mostly focused on the millimeter wave (mmwave) frequencies and in the context of 5G-NR. 

In this work, we extend the notion of wireless midhaul links and investigate the feasibility of deploying sub-THz and THz links for dense urban areas, which not only reduce the cost but also provide higher data rates. In particular, our work aims to unify the concepts of functional splits and THz communications to cater for the ever increasing demands on transport networks. We focus on maximizing the efficiency of wireless midhaul links at 140 GHz by provisioning the minimum number of CUs to serve multiple DUs while at the same time,  maximizing the data rate of each CU-DU link. 



\section{Problem formulation}
\label{sec:problem}
We consider the design of wireless midhaul links at 140 GHz in an urban environment, where we have $C$ central units (CUs), and $D$  distributed units (DUs). We further consider that each CU is equipped with $N$ antenna elements and each DU has $M$ antenna elements such that $M \leq N$. The path between a CU and DU can be either Line of Sight (LOS) or Non-LoS (NLOS), depending on the environment. 
In our scenario, the path $k$ between a CU and DU is parameterized by received power (\texttt{rx-power}), Angle of Departure (\texttt{AoD$(\theta, \phi)$}), CU location (\texttt{CU$_{XYZ}$}), DU location (\texttt{DU$_{XYZ}$}),   Angle of Arrival (\texttt{AoA$(\theta, \phi)$}), delay spread $\delta(k)$, and relative phase rotation $\alpha(k)$. 
Here, $\theta$ is the elevation angle and $\phi$ is the azimuth angle.

\begin{figure*}[!ht]
    \centering
    \includegraphics[width=0.87\textwidth]{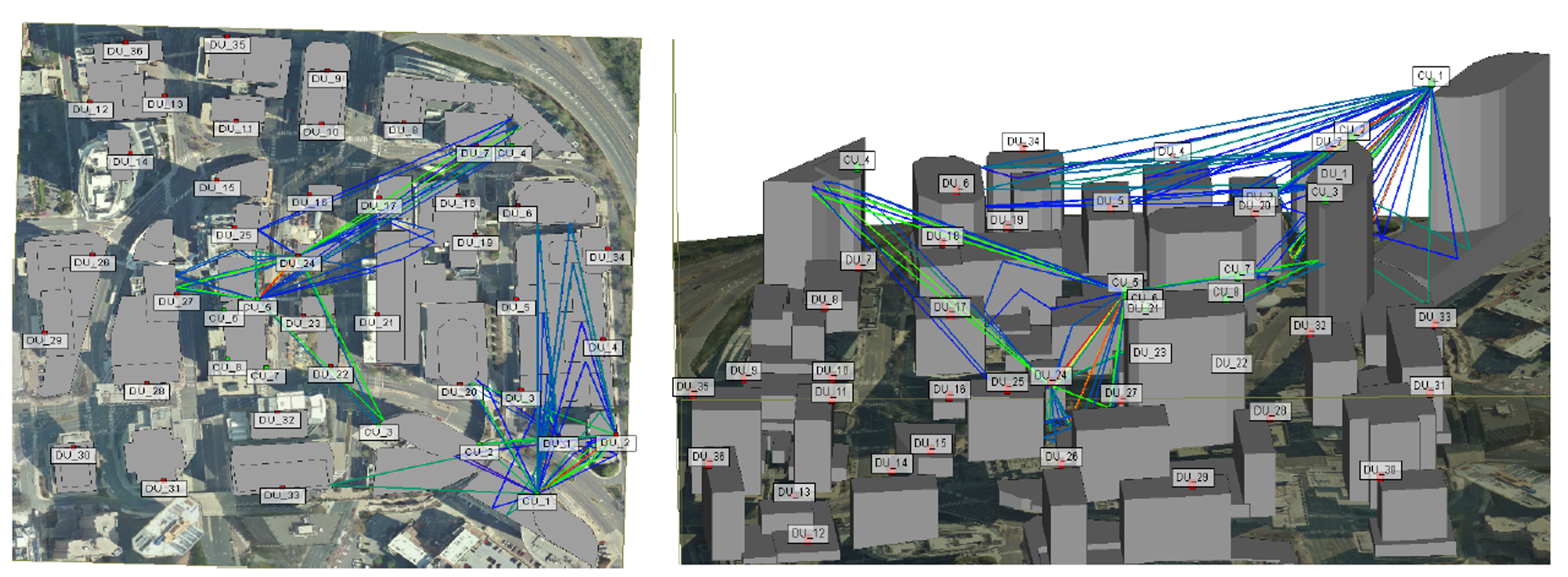}
    \caption{2D and 3D views of a $500 \times 500$ sq.m area in Rosslyn, VA, with 8 CUs and 36 DUs. A sample of propagation paths between the CUs and DUs are shown.}
    \label{fig:VA}
    \vspace{-3mm}
\end{figure*}

Based on this system model, our objective is to minimize the number of required CUs to serve all  DUs, while guaranteeing a peak rate $R$~\text{Gbps}, on each CU-DU link. Let us assume that $\mathcal{C}$ represents the set of CUs and $\mathcal{D}$ represents the set of DUs. Thus, we want to select a subset $\mathcal{C}^{'}$ of CUs that maximizes the effective coverage by associating multiple DUs.  Intuitively, we can consider the graph $\mathcal{G}=\{\mathcal{V},\mathcal{E}\}$, where $\mathcal{V}$ represents vertices of the graph as the CUs and DUs (i.e., $\mathcal{V}=\mathcal{C}\cup \mathcal{D}$), and $\mathcal{E}$ represents the link between the CUs and DUs. In this case, $\mathcal{E}$ includes both LOS and NLOS paths between the CUs and DUs. 

Let $X_i \in \{0,1\}$
be a binary variable that indicates whether CU $i$ belongs to  $\mathcal{C}^{'}$ or not, i.e., whether CU $i$ is selected to provide connectivity to DUs. Moreover, let $u_{ij} \in \{0,1\}$  be a binary variable indicating if a link in $\mathcal{E}$ from CU $i$ to DU $j$ carries DU $j$'s traffic. Furthermore, let the achieved link capacity between CU $i$ and DU $j$ is given by $r_{ij}$. Then, we can formulate the optimization problem for the selection of subset $\mathcal{C}^{'}$ as follows:     
\begin{equation}
\label{eqn:op2}
    \begin{cases}
\mathop{\mathrm{minimize}}\limits_{\{X_i\}} & \sum_{i \in \mathcal{C}} X_i \\
\text{subject to:} &  r_{ij} u_{ij} \geq R X_i, \ \forall i \in \mathcal{C}, \forall j \in \mathcal{D},  \\
&  \sum_{i=1 }^{C} u_{ij}  = 1, \ \forall j \in \mathcal{D}, \\
& u_{ij} = 1 - I\{X_i = 0\}, \forall i \in \mathcal{C}, \forall j \in \mathcal{D}. 
\end{cases}
\end{equation}

The objective function in \eqref{eqn:op2} is to minimize the number of required CUs to provide connectivity to all of the DUs. The first constraint guarantees that the link between the selected CU $i$ and DU $j$ provides the data rate of at least $R~\text{Gbps}$ only when the selected CU $i$ is provisioned i.e., only when $X_i=1$. The second constraint ensures that the CU $i$ and DU $j$ are associated only once, and the third constraint ensures $u_{ij}$ is zero when $X_i=0$. Note that the  capacity on each link also depends on  CU-DU association, CU-DU antenna alignment, channel estimation, and precoders. Next, we explain these steps.

\subsection{CU-DU association}
In general, in wireless networks the association between users and base stations is a fundamental and challenging problem due to various factors such as mobility~\cite{ashraf2016dynamic,randomWayPoint}. However, in our envisioned scenario, all the CUs and DUs are stationary and are fixed on rooftops of tall buildings,  as shown in Fig.~\ref{fig:VA}, where CUs are placed on taller buildings relative to DUs. Since we want to minimize the number of serving CUs, it is assumed that several DUs will be connected to each CU. The selection of CUs and association of CU-DU can be solved using a greedy algorithm as defined in Algorithm~\ref{alg:algo1}.
The first part of the algorithm is to select a subset of CUs $\mathcal{C}'$ for providing the data connectivity to all DUs and obtain a subgraph $\mathcal{G}'=\{\mathcal{V},\mathcal{E}'\}$ that can provide the required data rates to each DU. Let $\mathcal{D}''$ be the subset of DU's that are connected to a CU at each iteration and $\mathcal{D}'$ be the set of DUs that are already associated in the previous iterations. At each iteration, a CU is selected that has the maximum number of LOS paths to the DUs that were not previously associated with any of the CUs, ensuring that each DU is associated to a CU only once. After every iteration we update the sets $\mathcal{C}'$ and $\mathcal{D}'$ (line 4 and line 6).
The second part of the Algorithm~\ref{alg:algo1}  is to associate these new subset of CUs in $\mathcal{C}''$ to all the DUs in $\mathcal{D}$ and obtain a subgraph $\mathcal{G}'=\{\mathcal{V}',\mathcal{E}'\}$ where $\mathcal{V}'=\mathcal{C}' \cup \mathcal{D}$ and $\mathcal{E}'$ represents the links between CUs and DUs in the new subgraph $\mathcal{G}'$. In each iteration a subset of DUs $\mathcal{V}'$ are associated to a CU based on the highest signal strength received among the set $\mathcal{C}''$ (line 11) and the links between CU and DU $\mathcal{E}'$ are updated (line 12 and line 13).

\begin{algorithm}
 \caption{Greedy algorithm for CU selection and CU-DU association}
 \label{alg:algo1}
 \begin{algorithmic}[1]
 \REQUIRE Connectivity graph ($\mathcal{G}$) \\
 \ENSURE Data connectivity graph ($\mathcal{G}'$)  \\
 \STATE \textbf{Initialization:} $\mathcal{C}' \gets \phi$, $\mathcal{G}' \gets \phi$, $\mathcal{D}' \gets \phi$, $\mathcal{D}'' \gets \phi$,$\mathcal{V}' \gets \phi$, $\mathcal{E}'' \gets \phi$. \\ 
 \WHILE {$\mathcal{D}'\neq \mathcal{D}$}
 \STATE \textit{Select} node $v$ in $\mathcal{C} - \mathcal{C}'$ that satisfies the received signal power threshold to most of the nodes in $\mathcal{D} - \mathcal{D}'$. 
 \STATE $\mathcal{C}' = \mathcal{C}' \cup \{v\}$
 \STATE \textit{Include} set of DUs $\mathcal{D}''$ that are connected to $v$ in $\mathcal{D}'$.
 \STATE $\mathcal{D}' = \mathcal{D}' \cup \mathcal{D}''$
 \ENDWHILE
 \STATE $\mathcal{C}'' \gets \mathcal{C}'$
 \WHILE {$\mathcal{C}''$ is not empty.}
 \STATE \textit{Select} node $v$ in $\mathcal{C}''$. 
 \STATE \textit{Connect} DUs to CU $v$ with edge $E$ if it provides highest signal strength among CUs in set $\mathcal{C}''$
 \STATE $\mathcal{V}' = \mathcal{V}' \cup \{v\}$
 \STATE $\mathcal{E}' = \mathcal{E}'' \cup E$
 \STATE $\mathcal{G}' = \{\mathcal{V}',\mathcal{E}'\}$
 \STATE $\mathcal{C}'' = \mathcal{C}'' - \{v\}$
 \ENDWHILE
\STATE \RETURN Subgraph $\mathcal{G}'$ and $\mathcal{C}'$
\end{algorithmic}
\end{algorithm}


\subsection{CU and DU Beamforming}
      In the proposed system mode, CUs and DUs are assumed to have multiple antennas, thereby we can achieve beamforming gain. A simple antenna pattern with 3dB beam width of 65\textdegree ~for both horizontal and vertical radiation patterns is assumed for each antenna element whose specifications are provided in ~\cite{antennaspec}. The DU antenna array is aligned in the direction of strongest path. This can be easily computed from the captured AoA$(\theta,\phi)$. For CUs, however,  since there are a set of connected DUs, we approximately align the CUs boresight to  those connected DUs. This can be roughly  estimated by taking the average direction and then rotating the AoDs of those connected DUs along the mean direction computed. It should be noted that this method, while providing promising results, may not be optimal as some of the DUs may not be aligned in the boresight of the connected CU. Considering more complex beamforming schemes is postponed for our future works.
      

\subsection{Channel Estimation}
The exploitation of spatial diversity is the key idea used in space-division-multiple-access (SDMA) scheme, which significantly enhances the channel capacity. In our system model, the CUs and DUs at different locations exhibit different spatial signatures. After properly orienting the antenna arrays at both CU and DU, the spatial signatures $\textbf{G}_{DU}$ and $\textbf{G}_{CU}$ of the antenna arrays are computed. 
  Since each DU receives several paths, the channel matrix $\mathbf{H}$ can be assumed as a linear combination of all the possible paths and is computed as follows: 
\begin{equation}
\small
\begin{split}
\mathbf{H} = \sum_k \textbf{G}_{DU}*\textbf{G}_{CU}^T*\sqrt{g(k)}*e^{j2\pi (f* \delta(k)
 + \phi(k))}, 
\end{split}
\end{equation}
\normalsize
where $\textbf{G}_{DU}$ and $\textbf{G}_{CU}$ are the spatial signatures of DU and CU respectively, $g(k)$ is the effective gain on path $k$, $\delta(k)$ is the delay spread on the path $k$, and $\phi(k)$ is the relative phase observed on the path $k$. In our system configuration, most of the paths turned out to be LOS, hence we safely assumed the channel to be narrow-band while computing the channel matrix~$\mathbf{H}$.

\subsection{Capacity Calculation}
\begin{figure}
    \includegraphics[width=0.98\linewidth]{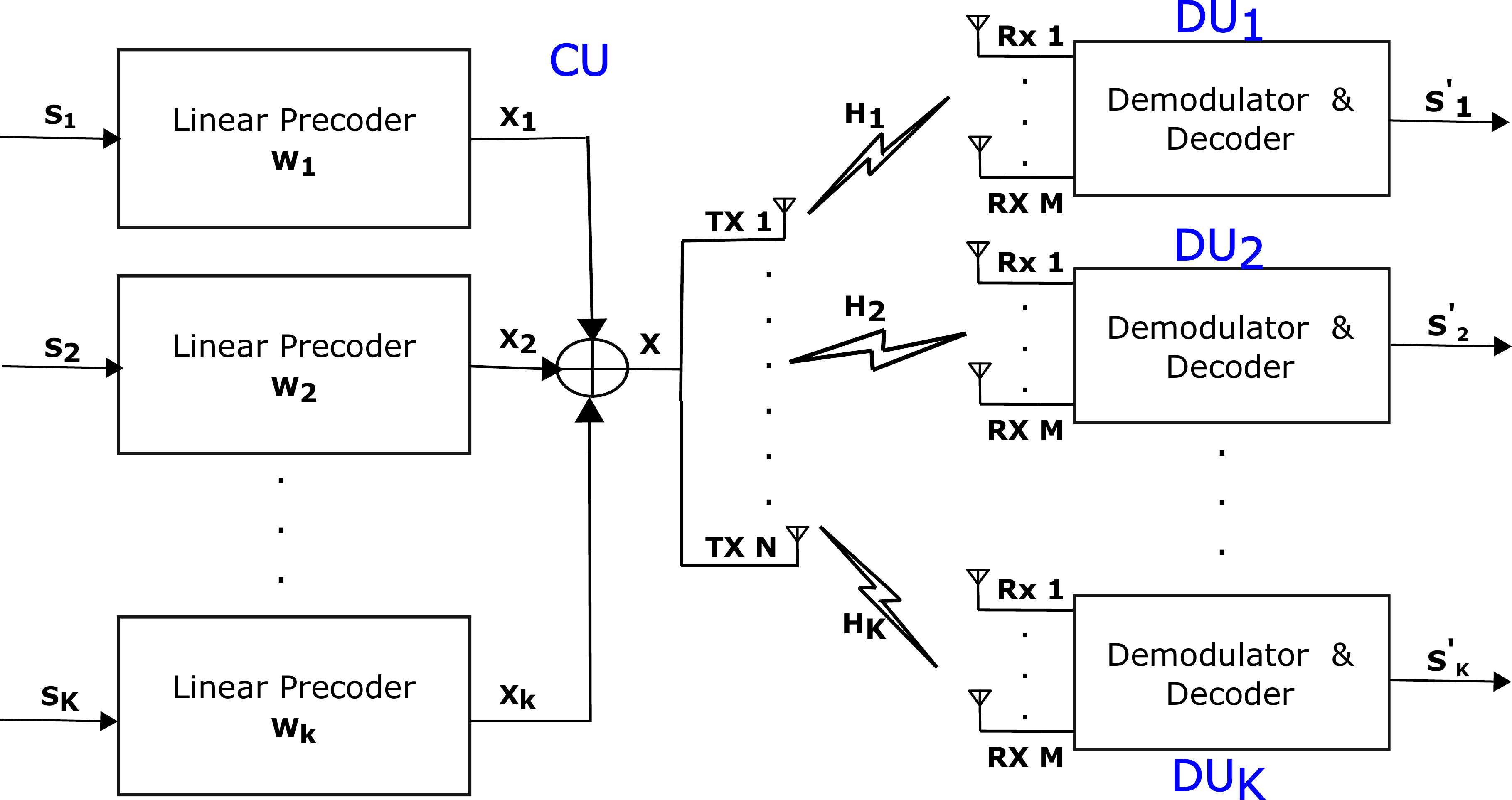}
    \caption{Downlink MU-MIMO configuration with linear precoding.}
    \label{fig:MUMIMO}
\end{figure}

Note that a single CU may provide connectivity to multiple DUs. As they operate on the same frequency simultaneously, we use SDMA and Multi User Multiple Input Multiple Output (MU-MIMO) for provisioning optimal link capacity between each CU-DU pair~\cite{heath2016overview}. Downlink MU-MIMO is a well-studied research problem~\cite{gesbert2007shifting}. Block diagram demonstrating the MU-MIMO with linear precoding is shown in Fig.~\ref{fig:MUMIMO}, where the CU has $N$ antenna elements, the DU has $M$ antenna elements, $S_k$ are the samples before precoding, $W_k$ is the precoding vector for samples $S_k$, $X_k$ is the precoded vector corresponding to $S_k$ all intended for the $k^{\text{th}}$ DU, $H_k$ is the channel observed by the $k^{\text{th}}$ DU, and $S_k ^{ '}$ is the decoded output corresponding to $S_k$.

In general, to maximize the downlink MU-MIMO system capacity, we require the maximization of Signal to  Interference plus Noise Ratio (SINR) using precoding techniques at the transmitter to suppress the co-channel interference (CCI). However, finding the optimal precoders that maximizes the SINR is computationally complex~\cite{sadek2007leakage,sadek2007active}. There are several schemes to perfectly cancel the CCI and in turn maximize the MU-MIMO capacity proposed in~\cite{spencer2004zero,bourdoux2002joint}, also known as Zero Forcing (ZF) techniques. However, they all impose a condition on the transmitter to possess sufficiently large number of antennas. In particular, the number of antennas at CU should be at least equal to the sum of antennas of all the connected DUs that makes the CU design more complicated. Alternatively, instead of optimizing the complex SINR problem, the scheme proposed in~\cite{sadek2007leakage,sadek2007active} uses a different approach based on signal leakage to compute the precoding vector that maximizes the Signal to Leakage plus Noise Ratio (SLNR). The SLNR is defined as:
\begin{equation}
\small
    \text{SLNR}_i = \frac{\norm{\textbf{H}_i \textbf{W}_i}^2}{M_i\sigma_i^2 + \norm{\tilde{\textbf{H}}_{i} \textbf{W}_i}^2 } ~,
\end{equation}
\normalsize
where,
\begin{equation}
\small
\tilde{\textbf{H}}_i = [\textbf{H}_1^T ...~\textbf{H}_{i-1}^T ~\textbf{H}_{i+1}^T~...~\textbf{H}_K^T]^T,  
\end{equation}
\normalsize
is an extended channel matrix that excludes $\textbf{H}_i$ (i.e., the channel matrix between the CU and $DU_i$), $\sigma_i^2$ is the noise variance and $M_i$ is the number of antenna elements at $DU_i$. From~\cite{sadek2007active}, the precoding vector $W_i$ that maximizes the SLNR is given by:
\begin{equation}
\small
    \textbf{W}_i\sim \max.~ \text{eigenvector}~\Bigg(\Big(M_i\sigma_i^2\textbf{I} + \tilde{\textbf{H}}_i^*\tilde{\textbf{H}}_i\Big)^{-1}~ \textbf{H}_i^*\textbf{H}_i \Bigg), 
\end{equation}
\normalsize
and the norm of $\textbf{W}_i$ is adjusted to unity.
Following the precoding techniques as shown above, and by allocating equal transmit power to all the connected DUs, we compute the rate at each DU as follows:     
\begin{equation}
\small
r = B(1-\beta)\times \min\left(\log_{2}(1+\Gamma),~\rho_{\max}\right), 
\end{equation}
\normalsize
where $r$ is the rate, $B$ is the bandwidth, $\Gamma$ is the SINR, $\beta$ is the loss factor that is assumed to be $0.15$, and $\rho_{max}$ is the maximum spectral efficiency per link that is assumed to be $5.9~\text{bits/s/Hz}$. This is a reasonable assumption for our beamforming scenario to transmit a single stream in one direction only. As suggested in \cite{alkhateeb2015limited,elbir2019hybrid}, more advanced beamforming techniques can achieve higher spectral efficiency and data rates per link.

\section{Simulation Results}
\label{sec:results}

\subsection{Ray-tracing Setup}
Conducting city scale measurements to evaluate wireless midhaul at $140$ GHz would require expensive hardware and man-power. Thanks to ray-tracing simulation, it is possible to characterize complex environments in 3D and obtain channel parameters like AoAs, AoDs, path gains, doppler spread, and delay on each path. We used the Remcomm Wireless Insite ray-tracer~\cite{Remcom}, which is widely used by the research community~\cite{li2015validation,alkhateeb2018deep}. The simulation setup used in Remcomm to capture the channel parameters can be found in Table~\ref{table:sim_setup}. The propagation model, $X3D$, used in Remcomm setup caters for the atmospheric absorption losses that are significant at such high frequencies~\cite{rappaport2019wireless}. The simulation was carried out using a default 3D model of a $500\times500$ sq.m area of Rosslyn, VA, as shown in Fig.~\ref{fig:VA}.
\begin{table}[h!]
\centering
\begin{tabular}{ |c|c| }
  \hline
  \textbf{Parameter} & \textbf{Value}\\ 
  \hline
  Number of transmitters (CUs)  & $8$ \\
  \hline
  Number of receivers (DUs) & $36$ \\
  \hline
   Carrier Frequency (f)& $140$ GHz \\
  \hline
   Bandwidth (B) & $2$ GHz \\
  \hline
  Transmit power& $43$ dBm \\
  \hline
  Transmit Antenna & $isotropic$ \\
  \hline
  Receive antenna & $isotropic$\\ 
  \hline
  Propagation model &$X3D$\\
  \hline
  Number of reflections &$6$\\  
  \hline
  Maximum number of paths &$25$\\ 
  \hline
 \end{tabular}
\caption{Ray-tracing simulation parameters}
\label{table:sim_setup}
\vspace{-3mm}
\end{table}

After selecting the study area, one can select propagation paths to be captured in the output. The tool then returns path parameters (rx-power, $AoD(\theta, \phi)$, $CU_{XYZ}$, $DU_{XYZ}$, $AoA(\theta, \phi)$, $delay$, and $phase$) in a text file for each simulated location, which are then fed into MATLAB for further processing. We assign a unique TX-ID and RX-ID to each path based on the CU and DU locations.

\subsection{MATLAB Post-processing Setup}
Since we assumed omni-directional antennas for both CU and DU in the Remcomm setup, and obtained the necessary path parameters, we can now apply MIMO channel characteristics in MATLAB. To this purpose, We use the MATLAB's phased array toolbox to generate antenna arrays for both CUs and DUs. All CUs and DUs are assumed to be $16\times16$ arrays and each antenna element is assumed to have an antenna pattern as specified by 3GPP~\cite{antennaspec}. 

\begin{figure}[h!]
    \includegraphics[width=\linewidth]{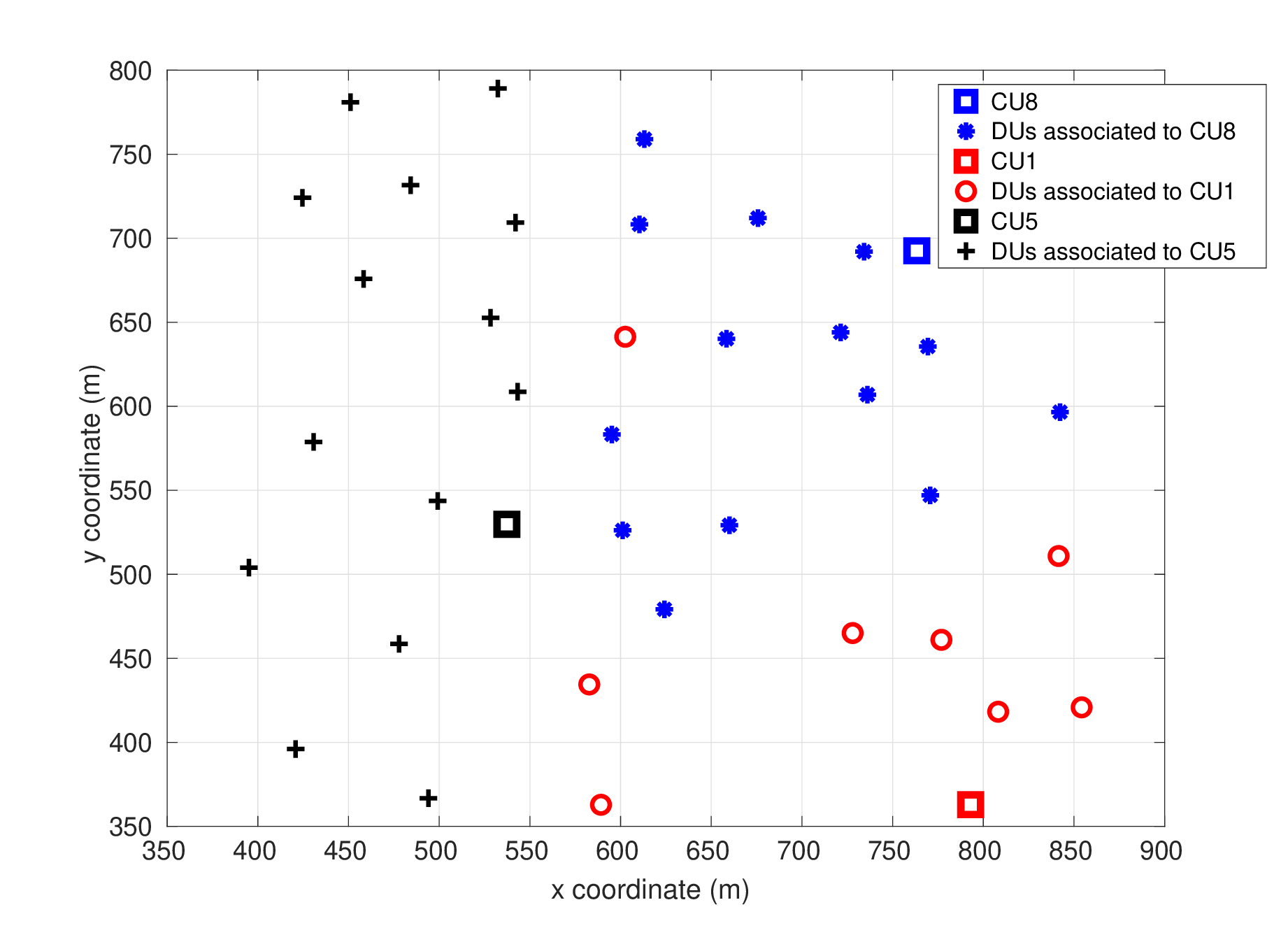}
     \caption{Initial association of CU-DU pairs with 3 CUs.}
    \label{fig:assoc}
\end{figure}

We start with CU-DU association as described in Algorithm~\ref{alg:algo1} by associating all the DUs to the selected CUs as shown in Fig.~\ref{fig:assoc}. An   Depending on the initial connectivity graph, the algorithm selects a subset of CUs. As an example, we show 3 CUs~($\text{CU}_8$, $\text{CU}_1$ and $\text{CU}_5$), and their associated DUs, color coded accordingly to differentiate their association. Initially, we start with a smaller set of CUs and if these CUs do not provide peak data rate to all DUs, we change the parameters of the Algorithm to obtain a larger set of CUs and then repeat the process. Since the optimization set involves only 8 CUs, we can also do an exhaustive search and try to associate all the DUs while maintaining the peak data rate on all the links. 

 After the initial CU-DU association, the next step is proper alignment of CU-DU based on the received $AoD$ and $AoA$ on each path as described in the previous section. Studies suggest that the data requirements on transport network requirements will be ever increasing, hence to account for higher performance metrics, our goal is to achieve a peak rate (R) of 10~Gbps on each link between a CU and their associated DUs, thereby maximizing the overall system capacity. Finally, to obtain the channel capacity we first calculate the channel matrix between each CU-DU pair. From the channel matrix, one can find channel capacity by adopting both SDMA and downlink MU-MIMO, which results in an efficient use of space, time and frequency to serve all the DUs at the same time.

\begin{figure}[h!]
\includegraphics[width=\linewidth]{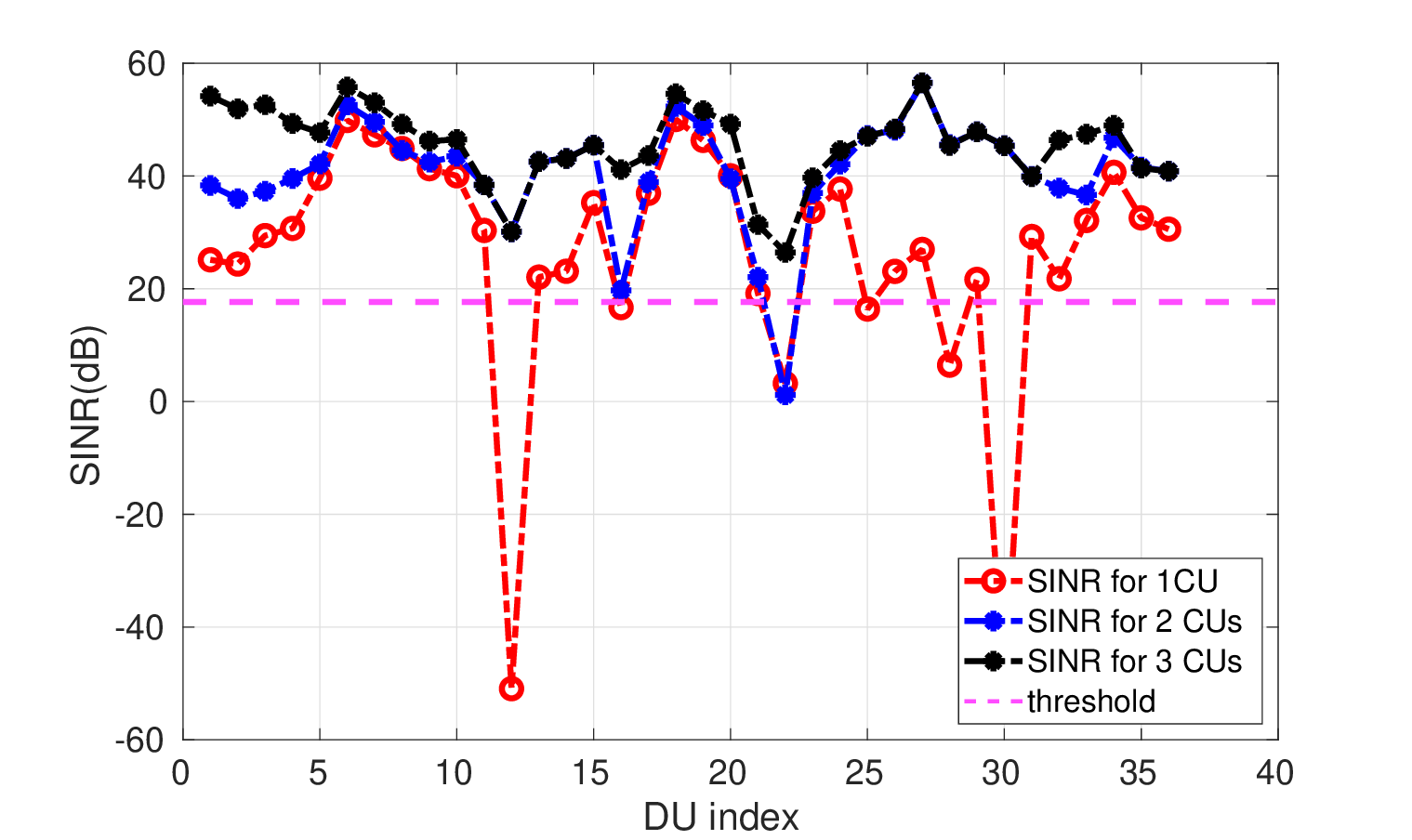}
\caption{SINR comparison with 1, 2, and 3 CUs with equal power allocation where each CU and DUs have 256 elements.  }
\label{fig:sinrs}
 \vspace{-3mm}
\end{figure}

\begin{figure}[h!]
\includegraphics[width=\linewidth]{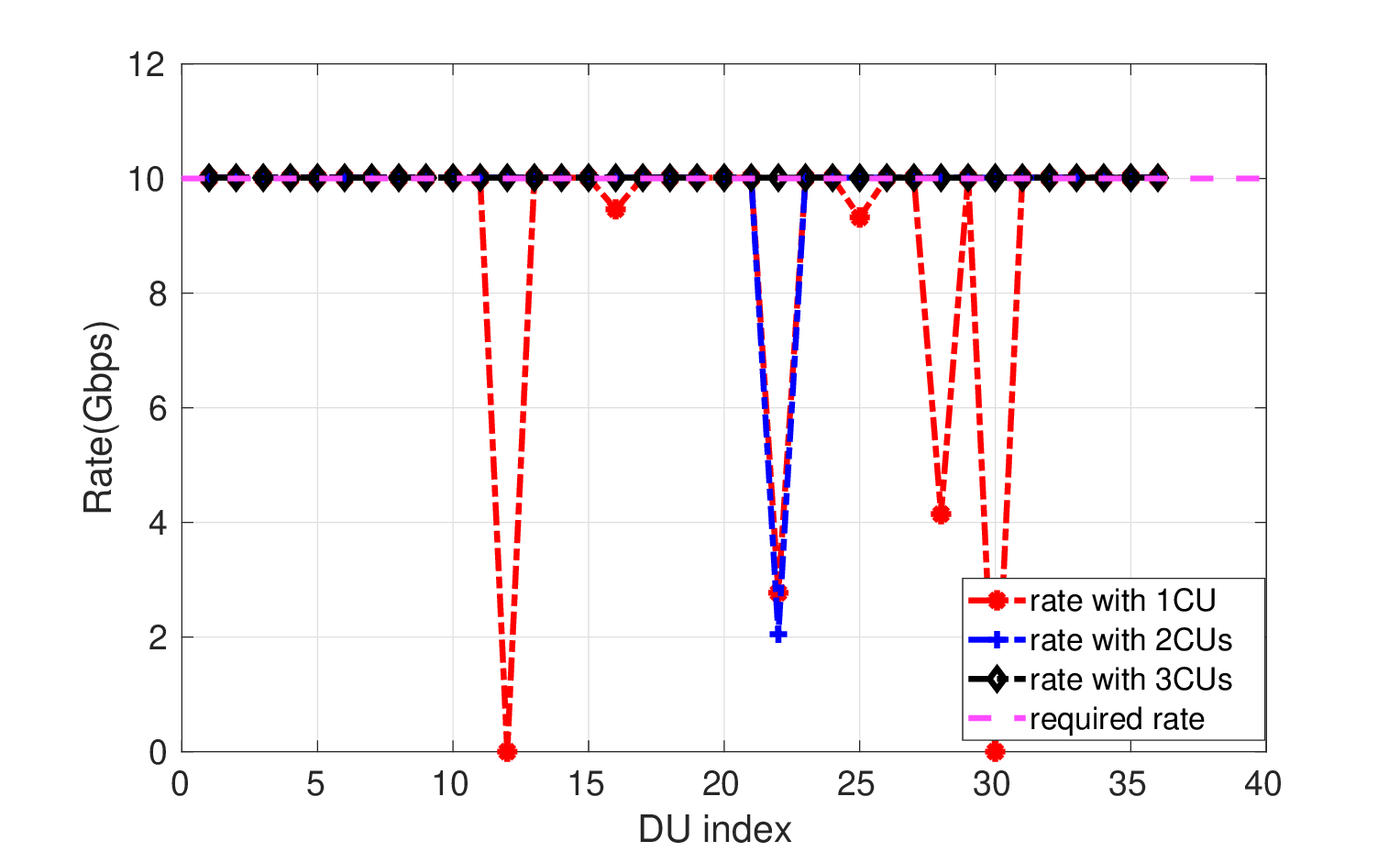}
\caption{Rates achieved with 1, 2, and 3 CUs with equal power allocation; all of the $36$ links achieve the desired rate with 3CUs}
\vspace{-3mm}
\label{fig:rates}
\end{figure}



\subsection{Effect of the Number of CUs and Antenna Elements}
Initially, we start with a smaller set of CUs (1 CU to start with) obtained by the greedy algorithm and associate all the DUs. Our simulation results demonstrated that with 1 or 2 CUs, it is not feasible to achieve peak rate on all the links. Also, we observed that when the CUs and DUs are assumed to have $16\times16$ arrays, the number of links that satisfy the peak rate constraint (R) in fact increases, compared with lower configurations (e.g., $8\times8$ or $16\times8$ arrays arrays).  As it can be seen in Fig.~\ref{fig:assoc}, we assume 3 CUs and associated all the DUs.
Fig.~\ref{fig:sinrs} shows the achieved SINR and Fig.~\ref{fig:rates} shows the data rates on all links. From the results, it can be seen that several links does not meet the required rate of $10$~Gbps when we consider 1 CU, and when we have 2 CUs only one DU (DU index 21) does not satify the peak rate constraint. We increase the number of CUs to 3 and again start by associating the DUs to the new set of CUs according to Algorithm~\ref{alg:algo1}. It can be noted that by adding a new CU all of the link rates operate at the peak rate.  In all the above scenarios, links with the SINR of at least $19.5~\text{dB}$ shown as dashed line in Fig.~\ref{fig:sinrs} and can achieve the peak rate of $10~\text{Gbps}$.

In our simulations we assumed that transmit power is equally allocated to all the connected DUs. In future, we will enhance this power allocation method by considering a joint optimization problem that extends Eq.~\eqref{eqn:op2}. A joint optimization problem would not only minimize the number of required CUs but also efficiently allocate power to the connected DUs such that the desired rate constraint is satisfied on all the links.
In our future study, we would also consider more advanced beamforming techniques (which can further extend our spectral efficiency assumption), clustering algorithms to make the DUs into separable groups, transmitting more than one stream of data, all of which would further improve the data rates. 

\section{Conclusion}
\label{sec:conclusion}

In this paper, we presented a preliminary assessment of midhaul links at 140 GHz. Using ray tracing, we evaluated wireless midhaul links in an urban scenario by considering an area of $500 \times 500$ sq.m. Our simulation results demonstrated that a set of 3 CUs can efficiently serve all of the 36 DUs. In fact, we achieve a peak rate of $10~\text{Gbps}$ on all of the links using an equal transmit power allocation.  Overall, it is evident that with the ever increasing traffic demands on transport networks new architectures are sought in the 5G-and-beyond era. Our results demonstrate that THz wireless midhaul links between CUs and DUs in conjunction with 3GPP functional splits is a promising direction especially for dense urban environments that merits further investigations.

\bibliographystyle{IEEEtran}
\bibliography{bibl}

\end{document}